\documentclass[10pt,aps,prd,twocolumn,superscriptaddress,amsmath, natbib, nofootinbib]{revtex4-1}
\usepackage{bm}
\usepackage{amsmath}
\usepackage{amsfonts}
\usepackage{aas_macros}
\usepackage{graphicx}
\usepackage[hidelinks]{hyperref}
\usepackage[capitalise, nameinlink]{cleveref}
\usepackage{booktabs}
\usepackage{color}
\graphicspath{{}}

\DeclareMathOperator{\sign}{sign}

\newcommand{\rad}{\mathrm{r}}
\newcommand{\mat}{\mathrm{m}}

\begin{document}

\title{Improved cosmological fits with quantized primordial power spectra}

\author{D. J. Bartlett}
\email{deaglan.bartlett@physics.ox.ac.uk}
\affiliation{Astrophysics Group, Cavendish Laboratory, J.J. Thomson Avenue, Cambridge, CB3 0HE, UK}
\affiliation{Trinity College, Trinity Street, Cambridge, CB2 1TQ, UK}
\affiliation{Astrophysics, University of Oxford, Denys Wilkinson Building, Keble Road, Oxford, OX1 3RH, UK}
\affiliation{Oriel College, Oriel Square, Oxford, OX1 4EW, UK}
\author{W. J. Handley}
\email{wh260@mrao.cam.ac.uk}
\affiliation{Astrophysics Group, Cavendish Laboratory, J.J. Thomson Avenue, Cambridge, CB3 0HE, UK}
\affiliation{Kavli Institute for Cosmology, Madingley Road, Cambridge, CB3 0HA, UK}
\affiliation{Gonville \& Caius College, Trinity Street, Cambridge, CB2 1TA, UK}

\author{A. N. Lasenby}
\email{a.n.lasenby@mrao.cam.ac.uk}
\affiliation{Astrophysics Group, Cavendish Laboratory, J.J. Thomson Avenue, Cambridge, CB3 0HE, UK}
\affiliation{Kavli Institute for Cosmology, Madingley Road, Cambridge, CB3 0HA, UK}

%%TC:ignore
\begin{abstract}
    We observationally examine cosmological models based on primordial power spectra with quantized wave vectors. Introducing a linearly quantized power spectrum with $k_0=3.225\times10^{-4}\mathrm{Mpc}^{-1}$ and spacing $\Delta k = 2.257 \times 10^{-4} \mathrm{Mpc}^{-1}$  provides a better fit to the \textit{Planck} 2018 observations than the concordance baseline, with $\Delta \chi^2 = -8.55$. Extending the results of \citet{RadPert}, we show that the requirement for perturbations to remain finite beyond the future conformal boundary in a universe containing dark matter and a cosmological constant results in a linearly quantized primordial power spectrum. It is found that the infrared cutoffs for this future conformal boundary quantized cosmology do not provide cosmic microwave background power spectra compatible with observations, but future theories may predict more observationally consistent quantized spectra.
\end{abstract}
%%TC:endignore

\maketitle

\section{\vspace{-10pt}\label{sec:Introduction}Introduction}
There has been much historical discussion of the significance of low-multipole features in the cosmic microwave background power spectrum~\cite{bennett2012,Planck_parameters_2013,2020A&A...641A...6P}, with many proposed primordial mechanisms to explain these features with varying degrees of naturalness~\cite{Chluba:2015bqa}.
In a recent paper, \citet{RadPert} proposed a novel mechanism for setting initial conditions on cosmological perturbations derived from considerations of the future conformal boundary in radiation dominated universes. The key prediction of this theory is a quantization of wave vectors for the primordial power spectrum of curvature perturbations. These quantized primordial spectra are capable of generating features in the cosmic microwave background (CMB) power spectrum such as suppression of power at low multipoles, features at intermediate multipoles and oscillations at higher multipoles.

\citet{RadPert} established in detail how such quantized spectra arise through consistency considerations of a linearized treatment of perturbations, and in particular their behaviour as they approach and then pass through the future conformal boundary (FCB). They also derived the evolution of particle gedoesics through the FCB, and discussed how one can interpret the nature of a universe beyond this point, and ``two-sheeted universes'' have been discussed by \citet{2021arXiv210906204B}. In \cref{sec:Physical Consequences} we fit these models to the latest \textit{Planck} 2018 CMB data~\cite{2020A&A...641A...5P,2020A&A...641A...8P} and find that whilst these models produce qualitatively interesting observational features they do not provide a better fit in comparison with the $\Lambda$CDM baseline.  In \cref{sec:Linear Quantization} we extend the scheme and scan through a class of parametrized quantized models, and find that some models within this class provide a markedly improved fit in comparison with the $\Lambda$CDM baseline. Such models are capable of reconstructing both the suppression of power in low-multipole CMB power spectra and the $20\lesssim\ell\lesssim30$ dip. We conclude in \cref{sec:conclusions} and discuss future extensions which may be able to predict {\it a priori\/} these best-fitting quantization schemes.

\vspace{-13pt}
\section{\label{sec:background}$\Lambda$CDM FCB quantization}
\vspace{-7pt}
In Ref.~\citep{RadPert} it was shown that for a universe with only radiation and a cosmological constant, the equations governing the background and perturbations are analytically continuable through the future conformal boundary. Making the requirement that the perturbative expansion remains finite throughout the domain results in there being only a discrete set of wave vectors that are allowed. 

\citet{RadPert} also showed that for a universe with only matter, the equations are analytically continuable if the energy density is viewed as $\rho_\mathrm{m}\sim a^{-3}=s^3$. However, as $s$ and $a$ are negative on the other side of the future conformal boundary, this results in radically different behaviour in contrast to the symmetry associated with the radiation-only case. As an alternative to this, one can define the material component as having a scaling $\rho_\mathrm{m}\sim |a|^{-3}$. This retains the symmetry on either side of the future conformal boundary, at the expense of analyticity since the modulus function $|s|$ is not differentiable at the FCB $s=0$. One can however ensure that things remain physical in spite of this by requiring that energy remains continuous for both the background and perturbations, which is shown to be equivalent to requiring that the solutions are symmetric or antisymmetric about the future conformal boundary. 

In this section, we consider universes with a cosmological constant alongside both noninteracting material and radiative components as an approximation to the concordance $\Lambda$CDM cosmology. As in Ref.~\citep{RadPert}, we start by approximating these components as perfect fluids and later consider the impact of higher-order multipoles in the Boltzmann hierarchy. However, unlike in Ref.~\citep{RadPert}, the equations for the combined case are not analytically solvable, so we must proceed using series expansion techniques.

\subsection{Background equations}
Throughout this paper we use units $8\pi G = c = \hbar = k_{\rm B} = 1$ and consider spatially flat universes in the Newtonian gauge with potentials $\Psi$, $\Phi$ such that the metric is defined by
\begin{equation}
    g_{\mu\nu}\mathrm{d}x^\mu \mathrm{d}x^\nu = a^2\left(\eta\right) \left[ \left(1 + 2\Psi\right) \mathrm{d}\eta^2 - \left(1 - 2\Phi\right) \mathrm{d}\vec{x}^2 \right].
    \label{eqn:metric}
\end{equation}
The background evolution of a universe filled with noninteracting barotropic fluids is defined by the Friedmann equation:
\begin{equation}
    H^2 = \frac{\dot{a}^2}{a^4} = \frac{1}{3}\sum_i \rho_i, \:\:\: \rho_i = 3H_0^2 \Omega_{i}|a|^{-3(1+w_i)},
    \label{eqn:friedmann}
\end{equation}
where $a$ is the scale factor, $\rho_i$ is the density of the $i$th fluid component with equation of state parameter $w_i$, $H_0$ is the present-day value of the Hubble parameter $H$, $\Omega_{i}$ is the fractional contribution of fluid $i$ to the universal energy budget today and derivatives with respect to conformal time are denoted with an overdot.

\subsection{Perturbation equations}

\subsubsection{Perfect fluid approximation}
Scalar perturbations $\delta\rho_i$ to the background densities $\rho_i$ are defined as $\delta_i \equiv \delta\rho_i / \bar{\rho}_i$, and the peculiar velocities of these fluid perturbations are defined to be $\vec{v}_i\equiv \vec{\nabla}v_i$.
To linear order, and assuming that all components are perfect fluids, the scalar perturbations evolve as~\cite{ma_bertschinger}
\begingroup
\allowdisplaybreaks
\begin{align}
    \dot{\Phi} &= - \frac{\dot{a}}{a}\Phi -\frac{1}{2}a^2 \sum_i (1+w_i)\rho_i v_i, 	\label{eqn:pert_phi}\\ 
    \dot{\delta}_i &= \left( 1 + w_i \right) \left(3 \dot{\Phi} + v_i k^2 \right) \label{eqn:pert_delta}, \\
    \dot{v}_i &= 3 \frac{\dot{a}}{a} \left(w_i - \frac{1}{3} \right) v_i - \left( \Phi + \frac{w_i}{1 + w_i} \delta_i \right). \label{eqn:pert_r}
\end{align}
\endgroup
One can also express potentials algebraically in terms of the other components via:
\begin{equation}
    \Phi = \Psi = \frac{1}{2k^2} \sum_i  \left( 3\frac{\dot{a}}{a} \left(1 + w_i \right) v_i  - \delta_i  \right) a^2\rho_i.
   \label{eqn:pert_phi_1}
\end{equation}
We take the fluid index $i$ to range over $i\in\{\rad,\mat,\Lambda\}$ for radiation $w_\rad=\frac{1}{3}$, matter $w_\mat=0$ and dark energy $w_\Lambda=-1$ respectively, but in general only perturb the first two of these.

It should be noted that \cref{eqn:metric,eqn:friedmann,eqn:pert_phi,eqn:pert_delta,eqn:pert_r,eqn:pert_phi_1} are symmetric under the transformation $a\to-a$, $\mathrm{d}\eta\to-\mathrm{d}\eta$, provided that the perturbations transform as $\Phi\to\pm\Phi$ and $\delta_i\to \pm\delta_i$, $v_i\to\mp v_i$. The opposing sign of $v_i$ is intuitive, since it is a velocity-like term that should change direction on changing the sign of $\mathrm{d}\eta$.

\subsubsection{Boltzmann hierarchy}

In reality one should describe photons via a distribution function in photon momentum and not as a perfect fluid. Decomposing the perturbation to this distribution function into momentum-averaged Legendre components, $F_{\rad\, \ell}$, and defining $G_{\rad\, \ell}$ to be the photon polarization component, we must solve the Boltzmann hierarchy~\cite{ma_bertschinger}
\begingroup
\allowdisplaybreaks
\begin{align}
&k^2 \Phi = - \frac{1}{2} a^2 \sum_i \left( \delta \rho_i + 3 \frac{\dot{a}}{a} \left( \bar{\rho}_i + \bar{P}_i \right) \frac{\theta_i}{k^2}  \right) \label{eqn:boltz_phi} \\ 
&k^2 \left( \Phi - \Psi \right) = \frac{3}{2} a^2  \sum_i \left( \bar{\rho}_i + \bar{P}_i \right) \sigma_i, \label{eqn:boltz_psi} \\ 
&\dot{\delta}_\mat = - \theta_\mat + 3 \dot{\Phi},  \label{eqn:boltz_dm} \\ 
&\dot{\theta}_\mat = - \frac{\dot{a}}{a} \theta_\mat + k^2 \Psi, \label{eqn:boltz_thm} \\ 
&\dot{\delta}_\rad = - \frac{4}{3} \theta_\rad + 4 \dot{\Phi}, \label{eqn:boltz_dr} \\ 
&\dot{\theta}_\rad = k^2 \left( \frac{\delta_\rad}{4} - \sigma_\rad \right) + k^2 \Psi + a n_{\rm e} \sigma_{\rm T} \left( \theta_{\rm b} - \theta_{\rad} \right), \label{eqn:boltz_thr} \\
\begin{split} \label{eqn:boltz_pi}
&\dot{F}_{\rad\, 2} = 2 \dot{\sigma}_\rad = \frac{8}{15} \theta_\rad - \frac{3}{5} k F_{\rad\, 3} - \frac{9}{5} a n_{\rm e} \sigma_{\rm T} \sigma_\rad \\
& \qquad + \frac{1}{10} a n_{\rm e} \sigma_{\rm T} \left( G_{\rad\, 0} + G_{\rad\, 2} \right),
\end{split} \\
\begin{split} \label{eqn:boltz_fl}
&\dot{F}_{\rad\, \ell} = \frac{k}{2 \ell + 1} \left[ \ell F_{\rad\, \left( \ell - 1 \right)} -  \left( \ell + 1 \right) F_{\rad\, \left( \ell + 1 \right)} \right]  \\
&\qquad - a n_{\rm e} \sigma_{\rm T} F_{\rad\, \ell}, \quad \ell \geq 3,
\end{split} \\
\begin{split} \label{eqn:boltz_gl}
& \dot{G}_{\rad\, \ell} = \frac{k}{2 \ell + 1} \left[ \ell G_{\rad\, \left( \ell - 1 \right)} -  \left( \ell + 1 \right) G_{\rad\, \left( \ell + 1 \right)} \right] \\
& \qquad + a n_{\rm e} \sigma_{\rm T} \bigg[ \frac{1}{2} \left( F_{\rad\, 2} + G_{r\, 0} + G_{\rad\, 2} \right) \left( \delta_{\ell 0} + \frac{\delta_{\ell 2}}{5} \right) \\
& \qquad  - G_{\rad\, \ell} \bigg],
\end{split}
\end{align}
\endgroup
where the subscript ${\rm b}$ refers to baryons, $n_{\rm e}$ is the electron number density, $\sigma_{\rm T}$ is the Thomson scattering cross section, and 
\begin{equation}
\theta_i = - k^2 v_i, \quad \sigma_i = \frac{\Pi_i}{6},
\end{equation}
for anisotropic stress $\Pi_i$. Once again, we consider the fluid index over the range ${i\in\{\rad,\mat,\Lambda\}}$ and perturb only the first two of these.

As in Ref.~\citep{RadPert}, we work in an approximation where, before recombination, there is tight coupling between matter and radiation, and therefore use the perfect fluid approximation. After recombination we assume free-streaming, and therefore set $n_{\rm e}=\sigma_\mat=0$. We see that this decouples $G_{\rad\,\ell}$ from the other perturbations, so we do not consider these terms further.   

\subsection{Initial conditions}
We may initialize the perturbation equations close to the singularity $\eta=0$ uniquely if we select the finite perturbative modes and consider only adiabatic perturbations so that $\delta_m = \frac{3}{4}\delta_r$ and $v_m = v_r$ as $\eta\to0$. Expanding \cref{eqn:friedmann,eqn:pert_delta,eqn:pert_r,eqn:pert_phi_1} as power series in $\eta$ under these constraints yields to first order:
\begingroup
\allowdisplaybreaks
\begin{align}
    a \left(\eta\right) &= H_0 \sqrt{\Omega_\rad} \eta + \mathcal{O}\left(\eta^2\right),\nonumber\\
    H \left(\eta\right) &= \frac{1}{H_0\sqrt{\Omega_\rad}}\frac{1}{\eta^2} - \frac{H_0\Omega_\mat^2}{16\Omega_\rad^{3/2}} + \frac{H_0^2 \Omega_\mat^3}{32 \Omega_\rad^2} \eta + \mathcal{O}\left(\eta^2\right),\nonumber\\
    \delta_\rad \left(\eta\right) &\propto -2 - \frac{H_0 \Omega_{m}}{4 \sqrt{\Omega_{r}}} \eta  +  \mathcal{O}\left(\eta^2\right),\label{eqn:ic_dr}\\
    v_\rad \left(\eta\right) &\propto - \frac{1}{2}\eta + \mathcal{O}\left(\eta^2\right),\label{eqn:ic_vr}\\
    \delta_\mat \left(\eta\right) &\propto -\frac{3}{2} - \frac{3H_0 \Omega_{m}}{16 \sqrt{\Omega_{r}}} \eta  +  \mathcal{O}\left(\eta^2\right),\label{eqn:ic_dm}\\
    v_\mat\left(\eta\right) &\propto - \frac{1}{2}\eta + \mathcal{O}\left(\eta^2\right),\label{eqn:ic_vm}\\
    \Phi\left(\eta\right) &\propto 1 - \frac{H_0 \Omega_{m}}{16 \sqrt{\Omega_{r}}} \eta +  \mathcal{O}\left(\eta^2\right),\label{eqn:ic_phi}
\end{align}
\endgroup
where the proportionality constant is the same for all expressions and without loss of generality following Ref.~\citep{RadPert} we define $\Phi=1$ at the singularity. Although written here to first order, we evaluate the velocity perturbations to fourth order, and the overdensity and potential to third order. We find that reducing this to third and second order, respectively, does not change our results.
	
\subsection{The future conformal boundary}

\begin{figure*}
\includegraphics{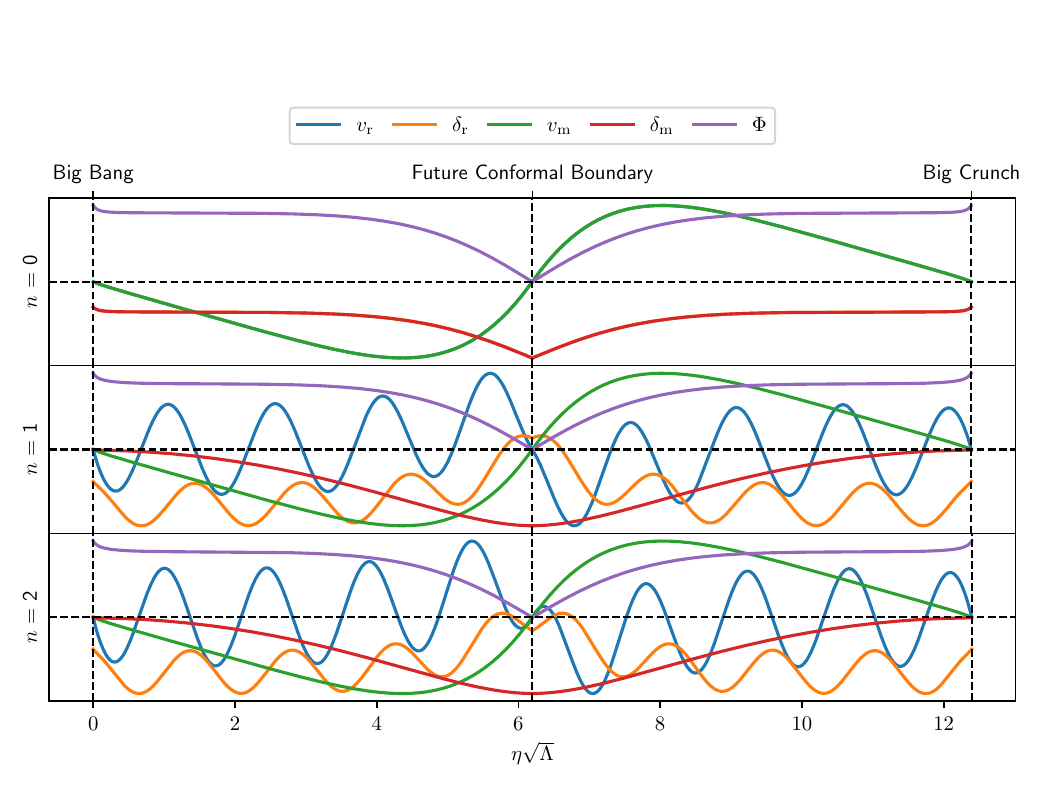}
\caption{\label{fig:first_allowed_pert} Radiation, matter and potential perturbations for the first three solutions which remain finite on both sides of the future conformal boundary in a $\Lambda$CDM universe containing only perfect fluids with cosmological parameters set to the \textit{Planck} 2018 best-fit values.}
\end{figure*}

We may now solve for the perturbations up to an arbitrary rescaling by integrating the background and perturbation equations beginning with the initial conditions \cref{eqn:ic_dr,eqn:ic_vr,eqn:ic_dm,eqn:ic_vm,eqn:ic_phi}. If we extrapolate this evolution beyond the present day, at some finite time in the conformal future we reach the future conformal boundary $\eta=\eta_\infty$. Although the scale factor diverges $a\to\infty$  as $\Delta\eta=\eta-\eta_\infty\to0_-$, all other terms in the background and perturbation equations remain finite, and the solution may be continued beyond this boundary. By symmetry, the background solution eventually arrives at a big crunch at $\eta=2\eta_\infty$. In order for our setup to be valid, following Ref.~\citep{RadPert} we should also demand that our perturbation variables remain finite at all times. For a general wave vector $k$, this will not be true, but for a discrete spectrum of wave numbers it can be.
As determined in Ref.~\citep{RadPert}, all of the technical considerations about perturbations remaining finite and analytic continuation through the future conformal boundary crystallize in practice into a symmetry requirement which we may numerically impose.

The conceptual strategy is therefore to compute the solutions of the perturbations at the future conformal boundary for each $k$, selecting those wave numbers whose solutions pass through the boundary with the correct symmetry (\cref{fig:first_allowed_pert}). 

\subsubsection{Perfect fluid approximation}

We start by considering perturbations for a perfect fluid, where the power series solutions about the future conformal boundary of \cref{eqn:friedmann,eqn:pert_phi,eqn:pert_r,eqn:pert_delta} take the form
\begin{widetext}
\begin{gather}
    \delta_\rad = \delta_\rad^\infty - \frac{2}{3k^2} \left[ \sign(\Delta\eta)\times 9H_\infty^3 \frac{\Omega_{\mat}}{\Omega_{\Lambda}}\left(\delta_\mat^\infty + 3 \dot{v}_\mat^\infty \right) - 2 \left(k^4 - 18 H_\infty^ 4 \frac{\Omega_{\rad}}{\Omega_{\Lambda}} \right)v_\rad^\infty \right] \Delta \eta + \mathcal{O}\left(\Delta\eta^2\right), \label{eqn:dr_expansion}\\
    v_\rad = v_\rad^\infty - \frac{1}{4} \delta_\rad^\infty \Delta\eta +\mathcal{O}\left(\Delta\eta^2\right), \label{eqn:vr_expansion}\\
    \delta_\mat = \delta_\mat^\infty - \frac{9}{2k^2}  \left[\sign(\Delta\eta)\times H_\infty^3 \frac{\Omega_{\mat}}{\Omega_{\Lambda}}\left(\delta_\mat^\infty + 3 \dot{v}_\mat^\infty \right)  +4H_\infty^4 \frac{\Omega_{\rad}}{\Omega_{\Lambda}} v_\rad^\infty \right] \Delta\eta + \mathcal{O}\left(\Delta\eta^2\right) \label{eqn:dm_expansion}\\
v_\mat = \dot{v}_\mat^\infty \Delta \eta + \mathcal{O}\left(\Delta\eta^2\right),\label{eqn:vm_expansion} \\
\Phi = -\frac{3}{2k^2}  \left[\sign(\Delta\eta)\times H_\infty^3 \frac{\Omega_{\mat}}{\Omega_{\Lambda}}\left(\delta_\mat^\infty + 3 \dot{v}_\mat^\infty \right)  +4H_\infty^4 \frac{\Omega_{\rad}}{\Omega_{\Lambda}} v_\rad^\infty \right] \Delta\eta + \mathcal{O}\left(\Delta\eta^3\right),\label{eqn:phi_expansion}
\end{gather}
\end{widetext}
where the Hubble constant at the future conformal boundary is calculated as $H_\infty = H_0\sqrt{\Omega_{\Lambda}}$. 

The four variables $\{v_r^\infty, \delta_\rad^\infty, \dot{v}_\mat^\infty, \delta_\mat^\infty\}$ are the leading terms in each solution's power series expansion, determine the higher-order terms in the series and are themselves determined up to an overall scaling by integrating \cref{eqn:friedmann,eqn:pert_phi,eqn:pert_r,eqn:pert_delta} up to $\eta_\infty$ beginning with the initial conditions \cref{eqn:ic_dr,eqn:ic_vr,eqn:ic_dm,eqn:ic_vm,eqn:ic_phi}, and can be viewed therefore as functions of $k$. For continuity, $\{v_\rad^\infty, \delta_\rad^\infty, \delta_\mat^\infty\}$ take the same value either side of the future conformal boundary; however this is not necessarily true for $\dot{v}_\mat^\infty$ as the derivative of $v_\mat$ could be discontinuous.

\cref{eqn:vr_expansion} is critical, as we can only have antisymmetry in $v_r$ if $v_r^\infty=0$. Assuming this is not true, we would require that the term depending on the sign of $\Delta\eta$ in \cref{eqn:dm_expansion} is the same on either side of the future conformal boundary, since this is added to a term proportional to $v_\rad^\infty$, which does not change sign. For this to be true, we require
\begin{equation}
	\delta_\mat^\infty  = - \frac{3}{2}  \left[ \dot{v}_\mat^\infty \left( \Delta \eta > 0 \right) +  \dot{v}_\mat^\infty \left( \Delta \eta < 0 \right) \right], {\rm \, if \,} v_\rad^\infty\neq 0.
\end{equation}
We find that we always have a finite $\delta_\mat$ at the future conformal boundary and therefore $\delta_\mat$ is required to be symmetric about this point. This is equivalent to requiring the linear term in \cref{eqn:dm_expansion} to change sign, hence
\begin{equation}
	8 H_\infty  \Omega_{\rad} v_\rad^\infty =  - 3 \Omega_{\mat} \left[ \dot{v}_\mat^\infty  \left( \Delta \eta > 0 \right) - \dot{v}_\mat^\infty \left( \Delta \eta < 0 \right) \right]
\end{equation}
Solving for $\dot{v}_\mat^\infty$ on either side of the future conformal boundary and substituting into \cref{eqn:dm_expansion}, we obtain
\begin{equation}
v_m = - \frac{1}{3} \left( \delta_\mat^\infty + {\rm sign}(\Delta \eta) \times 4 H_\infty \frac{\Omega_{\rad}}{\Omega_{\mat}} v_\rad^\infty \right) \Delta \eta  + \mathcal{O}\left(\Delta\eta^2\right).
\end{equation}
We have assumed $v_\rad^\infty\neq 0$ and used $\delta_\mat^\infty\neq 0$ but, if both of these are true, the above expression means we cannot make $v_\mat$ either symmetric or antisymmetric about the future conformal boundary and we thus arrive at a contradiction. The only allowed modes are those where $v_\rad^\infty = 0$. We note that in this case $\dot{v}_\mat^\infty$ is the same on either side of the future conformal boundary as the $v_\mat$ is antisymmetric about this point.

If no matter were present $(\Omega_{m}=0)$ then we could have either symmetry in $v_r$ if $\delta_r^\infty=0$ or antisymmetry if $v_r^\infty=0$. However, with the inclusion of matter, the coefficient of $\Delta\eta$ in \cref{eqn:dm_expansion} blocks one of these channels due to the presence of terms that depend on the sign of $\Delta\eta$. The choice $v_r^\infty=0$ automatically imposes symmetry on the remaining \cref{eqn:dr_expansion,eqn:dm_expansion,eqn:vm_expansion,eqn:phi_expansion}, so we have no further quantization conditions.

These series expansions also recover the results of Ref.~\citep{RadPert} if there is no radiation, since if $\Omega_{\rad}=0$ then \cref{eqn:dm_expansion,eqn:vm_expansion} have the correct symmetry however one chooses $\delta_\mat^\infty$ or $\dot{v}_\mat^\infty$. It should also be noted that at all points $\sign(\Delta\eta)$ is multiplied by $\Omega_{\mat}$, since it is the offending $|s|^3\ne s^3$ accompanying the material terms that prevents nonanalyticity in the expansions.

\subsubsection{Anisotropic stress}

As in Ref.~\citep{RadPert}, we now consider the properties of matter and radiation perturbations as they approach the future conformal boundary in the more realistic case where we do not treat radiation as a perfect fluid. To do this, we truncate \cref{eqn:boltz_phi,eqn:boltz_psi,eqn:boltz_dm,eqn:boltz_thm,eqn:boltz_dr,eqn:boltz_thr,eqn:boltz_pi,eqn:boltz_fl,eqn:boltz_gl} so we only retain terms with $\ell \leq 2$, so our only new term is $\Pi_\rad$. As before, we obtain power series about the future conformal boundary. \cref{eqn:dr_expansion,eqn:dm_expansion,eqn:vm_expansion} are unchanged with the addition of anisotropic stress. Our new expansions are
\begingroup
\allowdisplaybreaks
\begin{align}
v_\rad =& v_\rad^\infty - \frac{1}{12} \left( 3 \delta_\rad^\infty - 2 \Pi_\rad^\infty\right) \Delta\eta +\mathcal{O}\left(\Delta\eta^2\right), \label{eqn:vr_expansion_boltz}\\
\Pi_\rad =& \Pi_\rad^\infty - \frac{8}{5}k^2 v_\rad^\infty \Delta \eta + \mathcal{O}\left(\Delta\eta^2\right), \label{eqn:pi_expansion_boltz}\\
\begin{split}\label{eqn:phi_expansion_boltz}
\Phi =& -\frac{3}{2k^2}  \bigg[ \sign(\Delta\eta)\times H_\infty^3 \frac{\Omega_{\mat}}{\Omega_{\Lambda}}\left(\delta_\mat^\infty + 3 \dot{v}_\mat^\infty \right)\\
& +4H_\infty^4 \frac{\Omega_{\rad}}{\Omega_{\Lambda}} v_\rad^\infty \bigg] \Delta\eta + \mathcal{O}\left(\Delta\eta^2\right), 
\end{split} \\
\begin{split} \label{eqn:psi_expansion_boltz}
\Psi =& -\frac{3}{2k^2}  \bigg[ \sign(\Delta\eta)\times H_\infty^3 \frac{\Omega_{\mat}}{\Omega_{\Lambda}}\left(\delta_\mat^\infty + 3 \dot{v}_\mat^\infty \right)\\
& +4H_\infty^4 \frac{\Omega_{\rad}}{\Omega_{\Lambda}} v_\rad^\infty \bigg] \Delta\eta + \mathcal{O}\left(\Delta\eta^2\right),
\end{split}
\end{align}
\endgroup
which differ from before since the linear term of $v_\rad$ now contains a correction due to anisotropic stress and the potentials now contain a term $\mathcal{O} (\Delta\eta^2 )$, whereas before this was zero. It is at this order that $\Psi$ and $\Phi$ start to disagree. We have one more equation than before, and thus one more free parameter, $\Pi_\rad^\infty$.

Our arguments to enforce $v_\rad^\infty =0$ for a perfect fluid just considered the $\delta_\rad$, $\delta_\mat$ and $v_\mat$ perturbations, whose power series have not changed to linear order. Therefore, at this order in the Boltzmann hierarchy, our quantization condition is unchanged. As before, this assertion automatically enforces symmetry or antisymmetry in \cref{eqn:dr_expansion,eqn:dm_expansion,eqn:vm_expansion,eqn:vr_expansion_boltz,eqn:pi_expansion_boltz,eqn:phi_expansion_boltz,eqn:psi_expansion_boltz}, giving us a set of allowed wave numbers with a single condition.

\subsubsection{Higher-order corrections}
We previously neglected the term $F_{\rad\,3}$ in \cref{eqn:boltz_pi}, so we now reintroduce this. Denoting $F_{\rad\,\ell}^\infty$ as the value of $F_{\rad\,\ell}$ at the future conformal boundary, the expansion for $\Pi_\rad$ becomes
\begin{equation}
\Pi_\rad = \Pi_\rad^\infty - \frac{k}{5} \left( 8 k v_\rad^\infty + 9 F_{\rad\,3}^\infty  \right) \Delta \eta + \mathcal{O}\left(\Delta\eta^2\right),
\end{equation}
with all other series expansions unchanged at this order. Again, the quantization condition is $v_\rad^\infty=0$. However, $\Pi_\rad$ is no longer automatically either symmetric or antisymmetric, so we would need to introduce a second quantization condition
\begin{equation}
	\Pi_\rad^\infty = 0 {\quad \rm or \quad} F_{\rad\,3}^\infty = 0.
\end{equation}
We can generalize this to higher $\ell$ using \cref{eqn:boltz_fl}, obtaining a further condition for each $\ell > 3$
\begin{equation}
	\ell F_{\rad\,\left(\ell -1 \right)}^\infty = \left( \ell + 1 \right)  F_{\rad\,\left(\ell + 1 \right)}^\infty  {\quad \rm or \quad} F_{\rad\,\ell}^\infty = 0.
\end{equation}
There is no reason \textit{a priori} why this should occur for any one $k$ if we set each $F_{\rad\,\ell}$ to be zero at recombination, let alone for some set of these. Since $\{F_{\rad\,\ell}\}$ do not have to become nonzero at exactly the same conformal time, the freedom in $F_{r \ell}^\infty$ can be thought of as a freedom in when (near the surface of last scattering) $F_{r \ell}$ first becomes nonzero. By changing the latter the anisotropies at the FCB could be forced to have the correct symmetry.

A detailed analysis of how these initial conditions are related to each other is beyond the scope of this work, so we focus on $\ell \leq 2$ only. We refer to $\ell \leq 1$ and $\ell \leq 2$ as the ``Perfect Fluid'' and ``Imperfect Fluid'' cases respectively.

\section{\label{sec:Physical Consequences}Observational Consequences}

To compute the quantized power spectrum as detailed in the previous section we use the \textit{Planck} 2018 best-fit parameters from the \texttt{plik TTTEEE}+\texttt{lowl}+\texttt{lowE}+\texttt{lensing} likelihoods~\cite{2020A&A...641A...6P} (henceforth ``\textit{Planck} baseline'') for $H_0$, $\Omega_\Lambda$, $\Omega_{\mat}$ and $\Omega_\rad$. We integrate our background equations using \texttt{LSODA} from \texttt{scipy.integrate.solve{\_}ivp}~\cite{scipy} with relative and absolute precision of $10^{-13}$.
The resulting background is used to solve for perturbations up to recombination by integrating \cref{eqn:pert_phi,eqn:pert_r,eqn:pert_delta}. We set $\Phi=\Psi$ at recombination and let this equal the potential from the perfect fluid integration. Likewise, the density and velocity perturbations are assumed to be continuous at recombination and we set $\Pi_\rad = 0$ at this point. We now integrate \cref{eqn:boltz_phi,eqn:boltz_psi,eqn:boltz_dm,eqn:boltz_thm,eqn:boltz_dr,eqn:boltz_thr,eqn:boltz_pi} for $\ell \leq 2$ from recombination up to the future conformal boundary and find the perturbations which satisfy $v_\rad^\infty=0$ via a root-finding algorithm.

The first three allowed perturbations are plotted in \cref{fig:first_allowed_pert} for a perfect fluid. For numerical stability, we solve for $\log a$ instead of $a$ at small conformal times. 

\begin{figure}
    \includegraphics{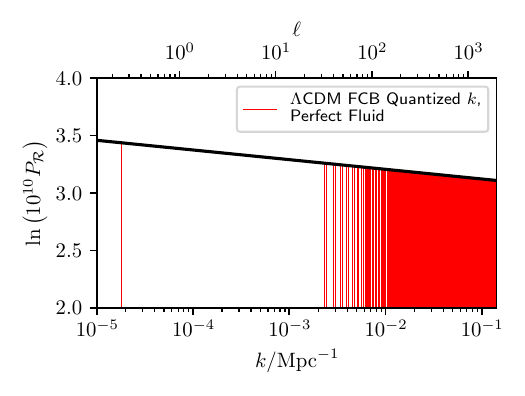}
    \caption{\label{fig:Quantized_Primordial}The quantized primordial power spectrum $P_\mathcal{R}\left(k\right)$. Red vertical lines indicate the first 500 allowed comoving wave numbers $k$ for a $\Lambda$CDM universe containing only perfect fluids with future conformal boundary quantization. Our quantization condition is that perturbations remain finite for all conformal times.}
\end{figure}

The quantized primordial power spectrum in the perfect fluid approximation is shown graphically\footnote{Throughout this paper we use an approximate conversion between $k$ and $\ell$ via the Limber approximation~\citep{2017JCAP...05..014L} $\ell \sim k D_{\rm A}$ where $D_{\rm A}$ is the comoving angular diameter distance to last scattering. This is a parameter-dependant conversion so for consistency we use the \textit{Planck} 2018 $\Lambda$CDM baseline cosmological parameters from~\citep{2020A&A...641A...6P} to define this transformation.} for the first 500 allowed wave numbers in \cref{fig:Quantized_Primordial}.

\begin{figure}
	\centering
    \includegraphics[width=\columnwidth]{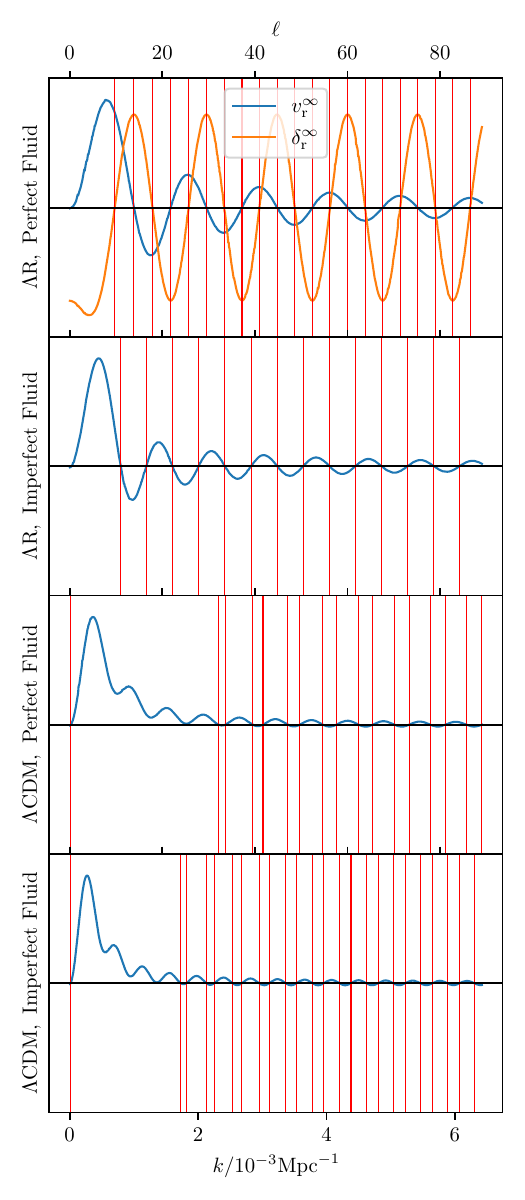}
    \caption{\label{fig:pert_at_fcb} Radiation velocity perturbation evaluated at the future conformal boundary, $v_\rad^\infty$, as a function of comoving wave number $k$ for a $\Lambda$R and a $\Lambda$CDM universe with \textit{Planck} baseline parameters. The perturbations associated with a given $k$ remain finite for all conformal times at the wave numbers shown with red lines. Note that the $\Lambda$R cosmology containing only perfect fluids does not have an allowed mode at the first zero of $v_r^\infty$~\citep{RadPert}. For this case we also plot the radiation density perturbation at the future conformal boundary, $\delta_r^\infty$, since in this case $\delta_\rad^\infty=0$ is also a permitted quantization condition.}
\end{figure}

In \cref{fig:pert_at_fcb} we plot the radiation velocity perturbation at the FCB and indicate the allowed wave numbers at the zeros of $v_\rad^\infty$. We see that the spacing is initially nonlinear, but, like in the pure radiation case, the wave numbers quickly settle down to linear. A linear fit to the allowed wave numbers $k / \sqrt{\Lambda} \leq 10$ yields a smallest two allowed wave numbers of
\begingroup
\allowdisplaybreaks
\begin{align}
k_0^{\rm perfect} &= 0.056 \sqrt{\Lambda} = 1.79 \times 10^{-5} {\rm \, Mpc^{-1}}, \\
k_1^{\rm perfect} &= 7.22 \sqrt{\Lambda} = 2.32 \times 10^{-3} {\rm \, Mpc^{-1}},
\end{align}
\endgroup
and asymptotic spacing
\begin{equation}
\Delta k^{\rm perfect} = 0.877 \sqrt{\Lambda} = 2.82 \times 10^{-4} {\rm \, Mpc^{-1}},
\end{equation}
for the perfect fluid, and
\begingroup
\allowdisplaybreaks
\begin{align}
k_0^{\rm imperfect} &= 0.042 \sqrt{\Lambda} = 1.34 \times 10^{-5} {\rm \, Mpc^{-1}}, \\
k_1^{\rm imperfect} &= 5.39 \sqrt{\Lambda} = 1.73 \times 10^{-3} {\rm \, Mpc^{-1}}, \\
\Delta k^{\rm imperfect} &= 0.657 \sqrt{\Lambda} = 2.11 \times 10^{-4} {\rm \, Mpc^{-1}},
\end{align}
\endgroup
when we include the effects of anisotropic stress.

As expected, since $v_r$ has an oscillatory complementary function near the FCB which varies as $\sim \cos \left( k \eta / \sqrt{3} \right)$, the spacing should be approximately $\Delta k = \sqrt{3}\pi/\eta_\mathrm{fcb} = 2.82 \times 10^{-4} {\rm \, Mpc^{-1}}$, which is consistent with the numerically calculated value of the perfect fluid case.

We note three differences in comparison with the pure radiation case:
\begin{enumerate}
\item The introduction of matter dramatically decreases the infrared cutoff since we do not have to exclude the $k/\sqrt{\Lambda} = \sqrt{2}$ mode.
\item The inclusion of matter creates ``missing'' modes; there are local minima in $v_{\rad}^{\infty}$ which do not have roots of $v_{\rad}^{\infty} = 0$ between them. This results in $k_1$ for $\Lambda$CDM universes being much larger than the first few allowed modes of $\Lambda$R universes. This is illustrated in \autoref{fig:first_allowed_pert}, where we see that the radiation density and velocity perturbations for the $n=1$ mode exhibit many cycles of oscillation before the FCB, although one would na\"{i}vely expect just one given the shape of the $n=0$ mode.
\item We only allow half the modes as in the perfect fluid radiation case (just the zeros of $v_r^\infty$ as opposed to both the zeros and turning points). However, since $\eta_{\rm FCB}$ is approximately twice as large in the $\Lambda$CDM case, the spacing is almost unchanged.
\end{enumerate}

Due to the linearity at high $n$, we will explicitly calculate the allowed wave numbers for $k / \sqrt{\Lambda} \leq 10$, then extrapolate to produce a complete spectrum.

To investigate the observational consequences of FCB quantization, we compute the predicted $C_\ell$ spectra using a modified version of \texttt{CLASS}~\cite{2011JCAP...07..034B}, changed to allow primordial power spectra with a discrete rather than continuous set of wave numbers (for more details, see the Appendix). We implement this by adapting parts of the code traditionally used for closed universes since these also have a quantized spectrum. We initially consider five cases:
\begin{enumerate}
    \item The traditional $\Lambda$CDM power spectrum.
    \item The quantized power spectrum arising from future conformal boundary considerations for a $\Lambda$CDM universe using the perfect fluid approximation.
    \item The quantized power spectrum arising from future conformal boundary considerations for a $\Lambda$CDM universe where we include the effects of anisotropic stress.
    \item The quantized power spectrum arising from future conformal boundary considerations for a $\Lambda$R universe from Ref.~\citep{RadPert}, using the perfect fluid approximation.
    \item The quantized power spectrum arising from future conformal boundary considerations for a $\Lambda$R universe from Ref.~\citep{RadPert}, where we include the effects of anisotropic stress.
\end{enumerate}

\begin{figure*}
    \includegraphics{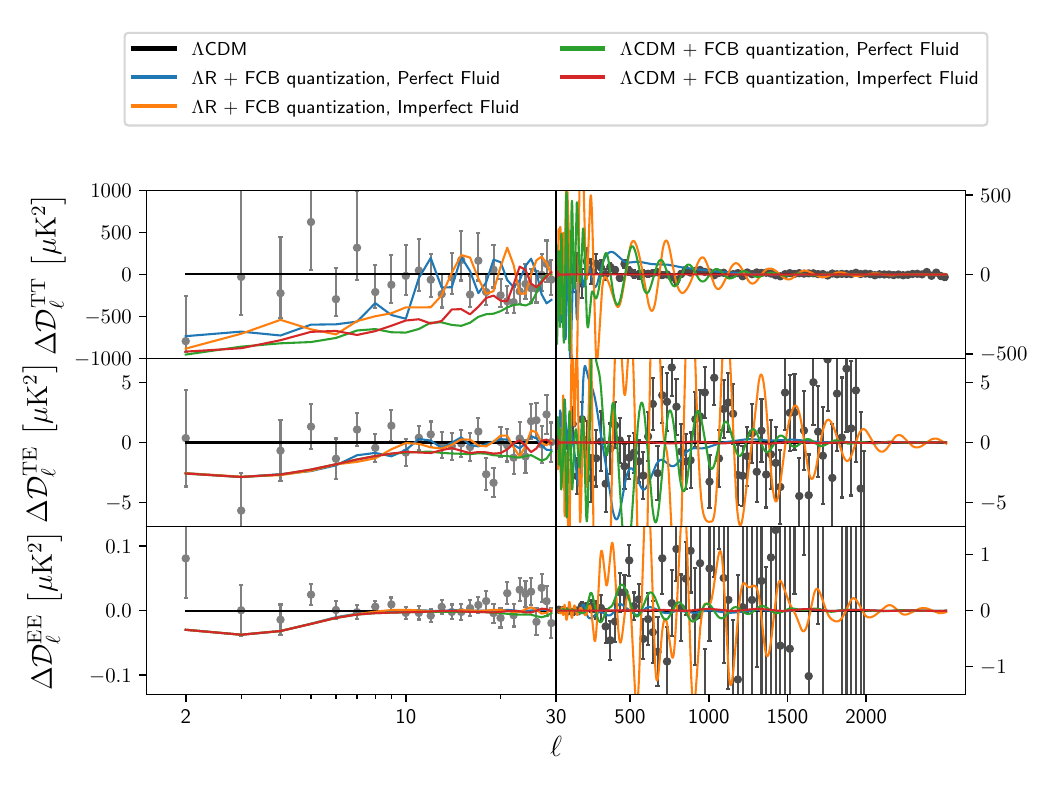}
    \caption{\label{fig:dl_diff_fcb} CMB power spectrum residuals between future conformal boundary quantized cosmologies and the $\Lambda$CDM baseline. $\Lambda$CDM + FCB quantization is in red and green, while $\Lambda$R + FCB is in blue and orange. The \textit{Planck} data residuals are also plotted for reference. The quantized cosmologies only allow comoving wave numbers $k$ such that perturbations remain finite for all conformal times. This restriction results in a drop in power at low $\ell$ due to a minimum allowed $k_0$ (and a corresponding rise at $k_0$) and oscillatory features from the finite spacing $\Delta k$ between allowed wave vectors. The cosmological parameters for all models are fixed to the \textit{Planck} baseline values.}
\end{figure*}

Assuming the \textit{Planck} baseline cosmological parameters, the power spectral differences between $\Lambda$CDM and the quantized cosmologies are plotted in \cref{fig:dl_diff_fcb} alongside the \textit{Planck} residuals.
There are two main features observed in all three plots. The first is a drop in power at low $\ell$, which is to be expected since there exists a minimum allowed $k$ in our quantized cosmology. We find $k_0$ corresponds to $\ell _0 < 1$ for $\Lambda$CDM FCB quantization, whereas $k_1$ corresponds to $\ell_1 =32$ for the perfect fluid, and $\ell_1 = 24$ for the imperfect case. This explains why we see smaller values of $\mathcal{D}_{\ell}$ for $\ell \lesssim 30$, but no rise in power at the smallest $\ell$, and thus the effective infrared cutoff can be considered to be $k_1$ in this scenario. The absence of the $k/\sqrt{\Lambda}=\sqrt{2}$ mode for a $\Lambda$R cosmology of perfect fluids explains the lack of the rise in that case and can therefore account for the low quadrupole~\cite{2015JCAP...06..014I} of the TT and octupole of the TE spectrum. The second disparity between the two cosmologies is an oscillatory behaviour at all $\ell$ values due to the spacing between allowed $k$ values. 

It is clear that these infrared cutoffs are implausible for both the $\Lambda$CDM and $\Lambda$R cases, and completely ruled out by modern cosmological observations. To test this, we use parameters from the \textit{Planck} posterior samples~\cite{Planck_legacy_archive} and find that we cannot obtain cosmological parameters which give a nonzero likelihood for the predicted CMB power spectra. We compute the first two allowed $k$ values for all parameters from the \textit{Planck} posterior samples, and find that the smallest value of $\ell_1$ is 30.5 for a perfect fluid, and 22.8 for the imperfect case. As will become apparent in \cref{sec:Linear Quantization}, these are too high to allow reasonable low-$\ell$ behaviour, although the $\Delta k$ obtained could be acceptable if more modes were allowed between $k_0$ and $k_1$. Since the imperfect fluid produced a smaller $k_1$ than the perfect fluid approximation, in future work we will investigate the effects of including higher-order terms in the Boltzmann hierarchy and more sophisticated modelling of recombination on the quantized primordial power spectrum, as this may reintroduce the ``missing'' modes in \autoref{fig:pert_at_fcb}.

\section{\label{sec:Linear Quantization}Linear Quantization in general}

Although FCB quantization does not provide quantitatively good fits to CMB power spectra, the previous section demonstrated that a quantized $k$ spectrum produces qualitatively interesting features such as a drop in power at low $\ell$ and a dip at $20\lesssim\ell\lesssim30$. Since the $\Lambda$R and $\Lambda$CDM FCB quantized spectra become linearly spaced at large $k$, we now consider whether a more general quantized, linearly spaced $k$ spectrum can provide a better fit to the \textit{Planck} data.

To do this, we introduce a finite spacing between allowed wave numbers $\Delta k$ and  a minimum allowed $k$ value $k_0$ such that
\begin{equation}
    k_n = k_0 + n \Delta k, \qquad n = 0, 1,2 \ldots
\end{equation}
where we treat $k_0$ and $\Delta k$ as free parameters.

\begin{figure*}
    \includegraphics{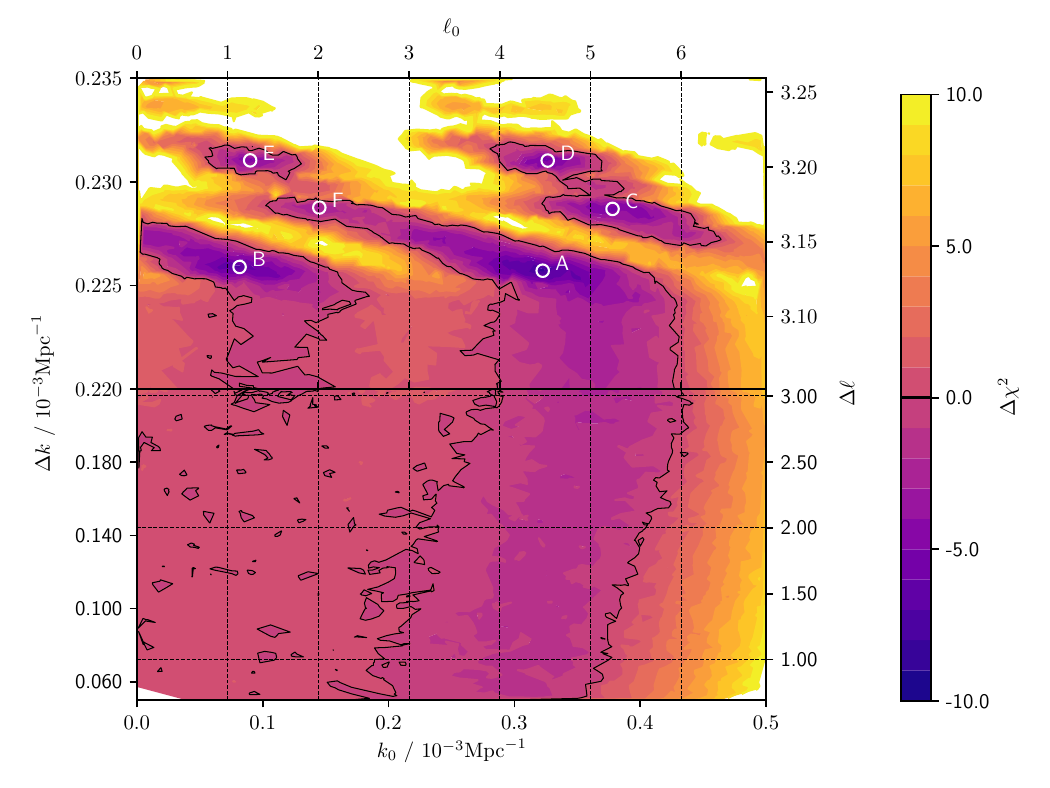}
    \includegraphics{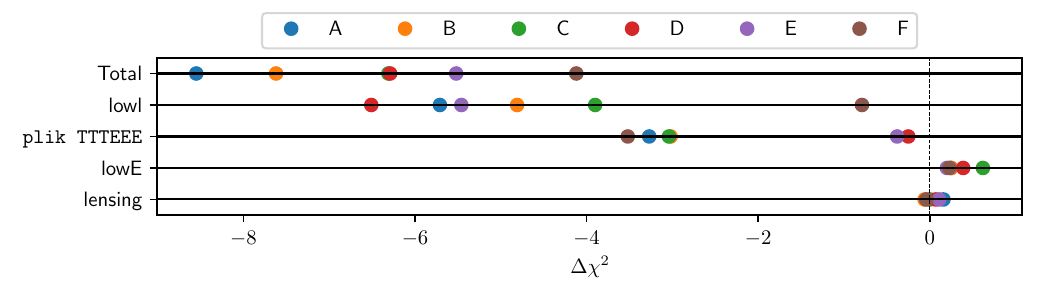}
    \caption{\label{fig:k0-deltak 2d} Upper: profile likelihood plot showing the difference of quality of fit $\Delta \chi^2$ between linearly quantized and $\Lambda$CDM models as a function of the first allowed comoving wave number $k_0$ and spacing $\Delta k$ optimized over all other cosmological and nuisance parameters. Negative $\Delta \chi^2$ indicates a better fit for the quantized model. 
        The best-fit point (A) has $\chi^2 = 2759.34$ which corresponds to $\Delta \chi^2 = -8.55$. 
        The plot was produced by exploiting the embarrassingly parallel nature of the problem by randomly choosing $\left(k_0, \Delta k\right)$ pairs, then optimizing over all other cosmological and nuisance parameters, which accounts for the small amount of noise in the contours.\\
        Lower: improvement in $\chi^2$ is driven by the reduction in \texttt{lowl}, although there is intriguingly also an improved high-$\ell$ contribution.
    }
\end{figure*}

The profile likelihood plot conditioned on $k_0$ and $\Delta k$ with the remaining parameters minimized over is shown in \cref{fig:k0-deltak 2d}. We optimize using the Nelder-Mead algorithm~\citep{Gao2012} with a simplex consisting of the parameters within the \textit{Planck} posterior samples with the highest likelihood.

For small $\Delta k$ (for which $\Delta \ell \lesssim 1$ and thus the spacing between $k$ values is negligible), we see that we can achieve a marginally better fit than the $\Lambda$CDM case if ${k_0 \sim 3 \times 10^{-4} \mathrm{Mpc}^{-3}}$. This is consistent with previous studies~\citep{Jing:1994jw,Sinha:2005mn} which found that introducing an infrared cutoff can slightly improve the low $\ell$ predictions.

The six labelled points (A-F) are local minima in $\chi^2$. The resulting $\chi^2$ and its contributions for these points are also shown in \cref{fig:k0-deltak 2d}. All these points have $\Delta \ell \gtrsim 3.1$, and thus we find that introducing a nontrivial finite spacing between $k$ values can significantly improve the value of $\chi^2$, by up to -8.55. As expected, this is driven by an improvement in \texttt{lowl}, although there is also a noticeable improvement in the high-$\ell$ likelihoods. The best-fit point (A) resides at
\begingroup
\allowdisplaybreaks
\begin{align}
    k_0 &= 3.225 \times 10^{-4} \, \mathrm{Mpc}^{-1} \\
    \Delta k &= 2.257 \times 10^{-4} \, \mathrm{Mpc}^{-1}.
\end{align} 
\endgroup
The cosmological and nuisance parameters which provide this best-fit are given in \cref{tab:parameters}, which are not significantly changed from the $\Lambda$CDM baseline~\citep{2020A&A...641A...6P}. The cosmological parameters are all consistent with those from the $\Lambda$CDM case, however the nuisance parameters shift a little more, but still at $\lesssim 2 \sigma$.

\begin{figure*}
    \includegraphics{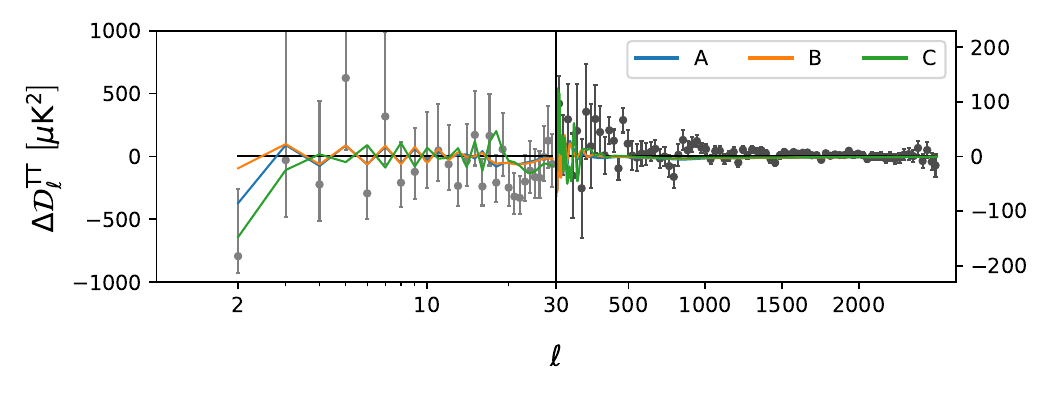}
    \caption{\label{fig:dl_diff_linear} CMB temperature residuals between the best-fitting linearly quantized cosmologies and the \textit{Planck} baseline $\Lambda$CDM model, fitted using the full \textit{Planck} baseline likelihood.
    Each of the lines correspond to the three best-fitting points from \cref{fig:k0-deltak 2d}. Our quantization condition introduces a drop in power at low $\ell$ and a dip at $20\lesssim\ell\lesssim30$.}
\end{figure*}

In \cref{fig:dl_diff_linear} we plot the CMB TT power spectra for points A-C. From this we identify the improvement as being due to a reduced quadrupole compared to the $\Lambda$CDM case and a dip following the \textit{Planck} data at $20\lesssim\ell\lesssim30$. The TE and EE spectra are not shown as the differences between the quantized and $\Lambda$CDM cases are negligible for these spectra.

\begin{figure*}
    \includegraphics{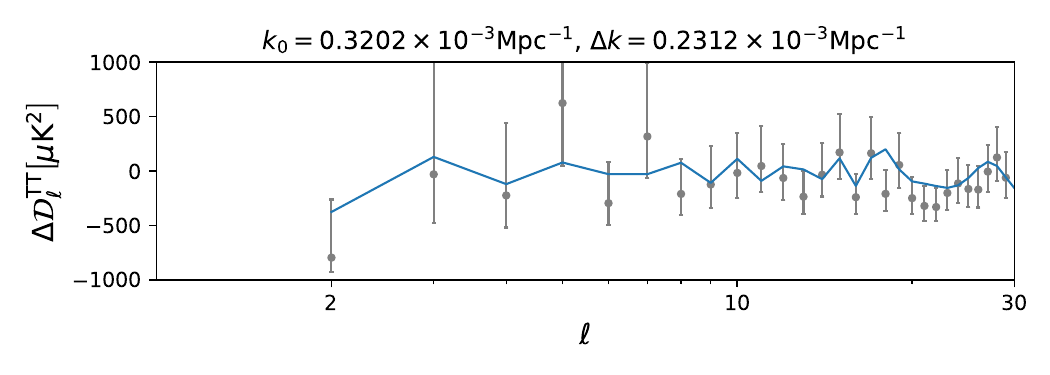}
    \caption{\label{fig:best_lowl} 
    CMB temperature residuals between the best-fit linearly quantized cosmology and the \textit{Planck} baseline $\Lambda$CDM model, fitted using just the \texttt{lowl} likelihood. Cosmological and nuisance parameters were fixed to the $\Lambda$CDM baseline. The best-fit $k_0$ and $\Delta k$ give $\Delta$\texttt{lowl}$=-6.97$. Without the influence of the high-$\ell$ likelihood, linearly quantized cosmologies are able to more accurately fit the $20\lesssim\ell\lesssim30$ dip.}
\end{figure*}
\begin{figure*}
    \includegraphics{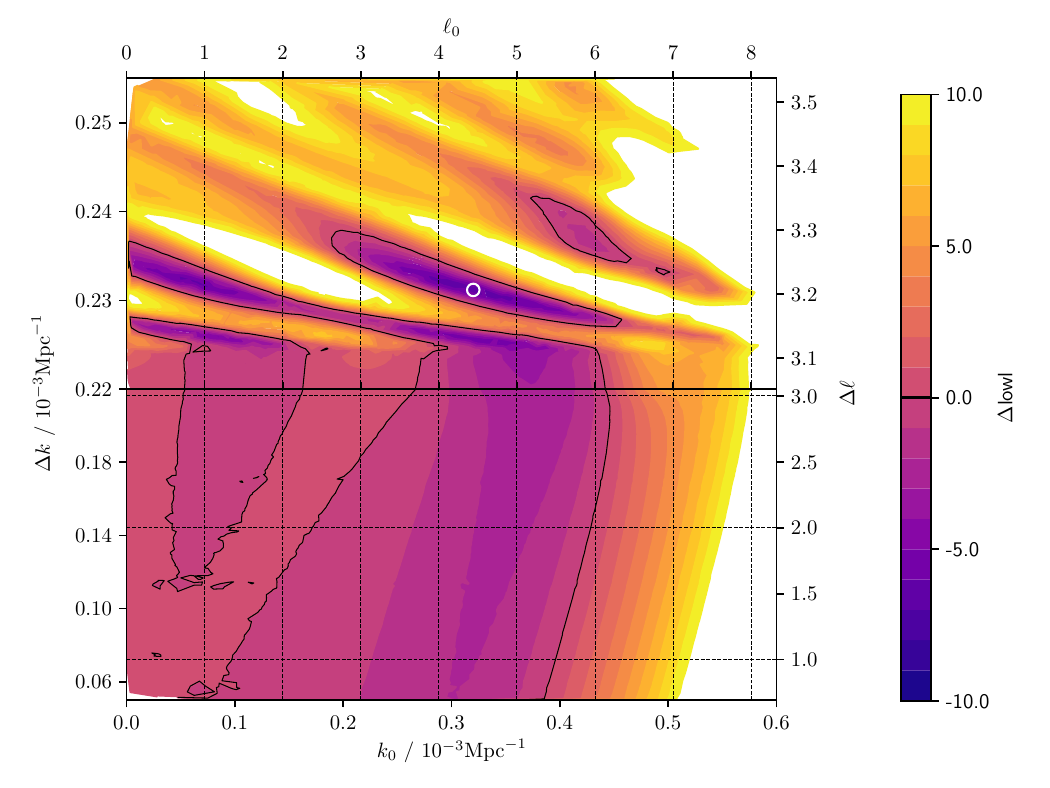}
    \caption{\label{fig:k0-deltak lowl} Difference of quality of fit $\Delta \texttt{lowl}$ between linearly quantized and $\Lambda$CDM models as a function of the first allowed comoving wave number $k_0$ and spacing $\Delta k$, where the cosmological and nuisance parameters are fixed to the $\Lambda$CDM baseline. Negative $\Delta$\texttt{lowl} indicates a better fit for the quantized model. 
        The best-fit point (circled) has $\texttt{lowl} = 16.57$ which corresponds to $\Delta \texttt{lowl}= -6.97$. 
    }
 \end{figure*}

The best-fit cutoff corresponds to $\ell_0 \sim 4.5$, compared to $\ell_1^{\rm (\Lambda CDM)} \sim 32$ and $\ell_0^{\rm (\Lambda R)} \sim 9.7$ for our FCB quantized models. Since the spacing $\Delta k$ is comparable for all our quantizations, we see that the reason for the poor fit in the future conformal boundary models is the large infrared cutoff. This resulted in a decrease in CMB power at too high a multipole, but from \cref{fig:dl_diff_linear} we see that a drop in power at $\ell\sim4.5$ provides a far more reasonable spectrum. 

Since \texttt{lowl} dominates the improvement in $\chi^2$, it is informative to consider optimizing over just \texttt{lowl}, fixing the remaining cosmological parameters to the \textit{Planck} baseline. In doing so we obtain the CMB TT power spectra plotted in \cref{fig:best_lowl}  and obtain a still stronger improvement of $\Delta\texttt{lowl}=-6.97$, with the improvement driven by an enhanced $20\lesssim\ell\lesssim30$ feature. We show how \texttt{lowl} varies with $k_0$ and $\Delta k$ in \cref{fig:k0-deltak lowl}.

\section{\label{sec:conclusions}Conclusions}

In this paper, in \cref{sec:background} we extended the results of Ref.~\citep{RadPert} to account for universes which simultaneously include radiation, cold dark matter and a cosmological constant, showing that both models produce similar quantized primordial power spectra. In \cref{sec:Physical Consequences} we examined the observational implications of these models, and found them to produce qualitatively interesting features in the cosmic microwave background power spectrum, namely a suppression of power in low-$\ell$, a dip at $20\lesssim\ell\lesssim30$ and oscillations at high-$\ell$. Quantitatively however, these models do not produce CMB power spectra consistent with modern cosmological observations. Inspired by the qualitatively interesting features generated by these quantized power spectrum models, in \cref{sec:Linear Quantization} we examined a wider class of linearly quantized models with quantized wave vectors $\{k_0 + n\Delta k:n=0,1,\ldots\}$. In this case we found that for values of 
${(k_0,\Delta k) = (3.225,2.257)\times 10^{-4} \mathrm{Mpc}^{-1}}$
these quantized primordial power spectra give markedly improved fits in comparison with the baseline concordance cosmological model, with $\Delta \chi^2=-8.55$. To a large extent this improvement is driven by the \texttt{lowl} likelihood, and when optimized only using \texttt{lowl} without the influence of high-$\ell$ likelihoods on the fit, one can extract a  $\Delta\texttt{lowl}=-6.97$.

It should be noted that whilst the results for linearly quantized primordial power spectra are interesting, far more work needs to be done before these can be considered feasible cosmological models. The $\Delta k$ parameter is highly fine-tuned, with the region of best-fit occupying roughly one percent of any reasonable prior. If these models were to be subject to a Bayesian analysis with evidences and parameter estimation~\cite{Trotta2008} there would be a subsequent Occam penalty which would penalize the quality of the fit. These results are also profiled, and sampling over the full parameter space would also degrade the quality of the fit. However, if there were models akin to the future conformal boundary quantizations that predicted {\em a priori\/} a quantized primordial power spectrum with $\Delta k \sim 2.257\times10^{-4}\,\mathrm{Mpc}^{-1}$, then these likelihood results suggest that these models would be candidates for a new concordance cosmology. In future work we will extend the future conformal boundary models to include higher-order multipoles in the Boltzmann hierarchy and a more sophisticated treatment of recombination to determine whether this class of models could achieve this.

There is also no need for these quantizations to be linearly spaced across the observable window. The fact that one can extract better low-$\ell$ fits by ignoring the high-$\ell$ likelihood suggests that if $\Delta k$ varied with $k$ still better fits might be obtained. There is also potential for using a free-form reconstruction approach~\cite{reconstructions} to determine the optimal locations and numbers of quantized wave vectors.

It is clear that there are many theoretical, observational and reconstructive investigations required into models with quantized primordial power spectra, but this work shows that there is compelling observational evidence for increased research effort into these cosmologies.

\begin{acknowledgments}
    We thank Metha Prathaban for useful discussions.
    
    D.~J.~B thanks the Cavendish Laboratory and Trinity College, Cambridge for their support during a Part III Project, is supported by STFC and Oriel College, Oxford, and acknowledges financial support from ERC Grant No 693024. W.~J.~H thanks Gonville \& Caius College for their support via a Research Fellowship and is supported by a Royal Society University Research Fellowship.

    This work was performed using resources provided by the \href{http://www.csd3.cam.ac.uk/}{Cambridge Service for Data Driven Discovery (CSD3)} operated by the University of Cambridge Research Computing Service, provided by Dell EMC and Intel using Tier-2 funding from the Engineering and Physical Sciences Research Council (capital grant EP/P020259/1), and \href{www.dirac.ac.uk}{DiRAC funding from the Science and Technology Facilities Council}.

    This work was based on observations obtained with \textit{Planck} (\url{http://www.esa.int/Planck}), an ESA science mission with instruments and contributions directly funded by ESA Member States, NASA, and Canada. 
\end{acknowledgments}

\appendix
\begin{table}
    \setlength{\heavyrulewidth}{0.5pt}
    \setlength{\abovetopsep}{4pt}
    \begin{ruledtabular}
        \begin{tabular*}{\columnwidth}{*3c}
            Parameter & Best-Fit Value & Change / $\sigma$ \\
            \toprule
            $\Omega_b h^2$ & 0.022279 & -0.36\\
            $\Omega_c h^2$ & 0.12031 & 0.04\\
            $100\theta_{\rm MC}$ & 1.041871 & 0.14\\
            $\ln \left(10^{10} A_s \right)$ & 3.0380 & -0.01\\
            $n_s$ & 0.96067 & -0.66\\
            $\tau$ & 0.0529 & 0.10\\
            \hline
            $y_{\text{cal}}$ & 1.00071 & 0.22\\
            $A^{CIB}_{217}$ & 45.5 & -0.20\\
            $\xi^{tSZ-CIB}$ & 0.60 & 0.86\\
            $A^{tSZ}$ & 8.69 & 0.58\\
            $A^{PS}_{100}$ & 195.7 & -1.23\\
            $A^{PS}_{143}$ & 28.0 & -1.00\\
            $A^{PS}_{143\times217}$ & 35.8 & -0.18\\
            $A^{PS}_{217}$ & 106.2 & -0.60\\
            $A^{kSZ}$ & 9.763 & 2.08\\
            $A^{{\rm dust}TT}_{100}$ & 7.65 & 0.08\\
            $A^{{\rm dust}TT}_{143}$ & 17.00 & 1.66\\
            $A^{{\rm dust}TT}_{143\times217}$ & 24.83 & 1.03\\
            $A^{{\rm dust}TT}_{217}$ & 99.1 & 0.37\\
            $A^{{\rm dust}TE}_{100}$ & 0.02 & -0.95\\
            $A^{{\rm dust}TE}_{100\times143}$ & 0.12 & 0.26\\
            $A^{{\rm dust}TE}_{100\times217}$ & 0.53 & -0.25\\
            $A^{{\rm dust}TE}_{143}$ & 0.279 & 1.00\\
            $A^{{\rm dust}TE}_{143\times217}$ & 0.66 & -0.17\\
            $A^{{\rm dust}TE}_{217}$ & 2.202 & -0.20\\
            $c_{100}$ & 0.99689 & -1.90\\
            $c_{217}$ & 0.99738 & -0.81\\
            \hline
            $k_0 / 10^{-3} \mathrm{Mpc}^{-1}$ & 0.3225 & - \\
            $\Delta k / 10^{-3} \mathrm{Mpc}^{-1}$ & 0.2257 & - \\
        \end{tabular*}
    \end{ruledtabular}
    \caption{\label{tab:parameters}Cosmological and nuisance parameters which optimize  $\chi^2$ for a linearly quantized $k$ spectrum. The parameter shifts between the quantized spectrum and the $\Lambda$CDM baseline are given in terms of the posterior parameter widths $\sigma$~\citep{2020A&A...641A...6P}.}
\end{table}

\section{\label{sec:Definitions Appendix}Power spectra with quantized $k$}

In this appendix we follow the derivations given in Ref.~\citep{Ma:1995ey}, but adapted for a quantized spectrum. 

Consider some function $f\left( \vec{x}, \hat{n}, \eta \right)$. Since a continuous spectrum of comoving wave numbers $k=|\vec{k}|$ is usually assumed in perturbation analysis, the canonical definitions for various quantities of interest are defined in terms of integrals. We must therefore rewrite our fields as Fourier series, defined as
\begin{equation}
    f\left( \vec{x}, \hat{n}, \eta \right) = \sum_k k^2 \Delta k  \int \mathrm{d}\Omega_k   e^{i \vec{k}\cdot\vec{x}} f\left(\vec{k}, \hat{n}, \eta\right),
\end{equation}
where $\Omega_k$ denotes the solid angle in $k$ space, we are looking in the direction of the unit vector $\hat{n}$ and where $\eta$ is conformal time
$\mathrm{d}\eta \equiv a \mathrm{d}t$
for scale factor $a\left(\eta\right)$ with cosmic time $t$.

We introduced the spacing $\Delta k \left( k \right)$ between allowed wave numbers $k=|\vec{k}|$ as a weighting to our Fourier coefficients so that all variables have the same dimensions as in the continuous case.  

At the origin ($\vec{x}=\vec{0}$),
\begin{equation}
    f\left(\hat{n}\right) = \sum_{\ell=0}^\infty \sum_{m=-\ell}^\ell a_{\ell m} Y_{\ell m} \left(\hat{n}\right),
\end{equation}
where, by orthogonality,
\begin{equation}
    a_{\ell m} = \int \mathrm{d}\Omega f\left(\hat{n}\right) Y_{\ell m}^* \left(\hat{n}\right).
\end{equation}

Expanding in term of Legendre polynomials,
\begin{equation}
\begin{split}
    f \left( \vec{x}, \hat{n}, \tau \right) = \sum_k  &k^2 \Delta k \int \mathrm{d}\Omega_k \sum_{\ell=0}^\infty \left(-i\right)^\ell \left( 2\ell + 1\right)  e^{i \vec{k}\cdot\vec{x}} \\
    & f_\ell\left(\vec{k}, \tau \right) P_\ell \left( \hat{k}\cdot\hat{n}\right),
\end{split}
\end{equation}
and using the identity
\begin{equation}
    P_{\ell^\prime}\left( \hat{k}\cdot\hat{n}\right) = \frac{4\pi}{2\ell^\prime +1} \sum_{m^\prime} Y_{\ell^\prime m^\prime}^* \left(\hat{k}\right) Y_{\ell^\prime m^\prime} \left(\hat{n}\right),
\end{equation}
we obtain
\begin{equation}
    a_{\ell m} = \left(-i\right)^\ell 4\pi \sum_k k^2 \Delta k \int \mathrm{d}\Omega_k Y_{\ell m}^*\left(\hat{k}\right) f_\ell\left(\vec{k}, \tau\right).
\end{equation}

We now define
\begin{equation}
    f_\ell\left(\vec{k}, \tau\right) \equiv \psi_i\left(\vec{k}\right)f_\ell\left(k, \tau\right),
\end{equation}
since the evolution equations are independent of $\hat{k}$, where $\psi_i\left(\vec{k}\right)$ is the initial perturbation and $f_\ell \left( k, \tau \right)$ is the photon transfer function in the case $f \left( \vec{x}, \hat{n}, \tau \right) =\Delta \left( \vec{x}, \hat{n}, \tau \right) $. 

We define
\begin{equation}
    \left< \psi_i \left(\vec{k}\right) \psi_i\left(\vec{k}^\prime \right) \right> \equiv P_\psi\left( k\right) \frac{\delta^{(K)}_{k k^\prime}}{k^2 \Delta k} \delta^{(D)}\left(\hat{k} + \hat{k}^\prime\right),
\end{equation}
where $\delta^{(K)}_{ab}$ is the Kronecker delta, and $\delta^{(D)}(\vec{a} - \vec{b})$ is the Dirac delta function. As before, the $\Delta k$ is introduced in the definition to be dimensionally consistent with the continuous case. We also define the power spectrum to be
\begin{equation}
    \left< a_{\ell m} a^*_{\ell^\prime m^\prime} \right> \equiv C_\ell \delta^{(K)}_{\ell \ell^\prime} \delta^{(K)}_{m m^\prime},
\end{equation}
and therefore
\begin{equation}
    C_\ell = \left(4\pi\right)^2 \sum_k k^2 \Delta k P_\psi\left(k\right) f_\ell^2\left(k, \tau\right).
\end{equation}

Consequently, we can summarize the continuous to quantized crossover as ``replace by a sum'', where we mean
\begin{equation}
    \int (\cdots)\:\mathrm{d} k  \to \sum_k (\cdots)\Delta k ,
\end{equation}
when we wish to convert an expression defined with a continuous $k$ spectrum to one with a quantized spectrum.

As usual, $C_\ell$ is related to $\mathcal{D}_\ell$ as
\begin{equation}
    \mathcal{D}_\ell \equiv \frac{\ell\left(\ell+1\right)}{2\pi} C_\ell.
\end{equation}

Since the basis $e^{i \vec{k} \cdot \vec{x}}$ is orthogonal irrespective of whether we do an integral or sum, we solve the same equations for perturbations as in the continuous case. 

One caveat of introducing the summation instead of an integral is that, when deriving the curvature perturbations from inflation, we cannot use a completeness relation when quantizing the inflaton. Instead of dwelling on this, we assume an initially (nearly) scale-invariant power spectrum, 
\begin{equation} \label{eqn:Primordial Power Spectrum}
    \mathcal{P}_\mathcal{R} \left( k \right) = A_s k^{n_s -1},
\end{equation}
without concern over its origin.

\bibliographystyle{unsrtnat}
\bibliography{my_references}

\end{document}